\documentclass[fleqn,usenatbib]{mnras}

\usepackage{newtxtext,newtxmath}

\usepackage[T1]{fontenc}

\DeclareRobustCommand{\VAN}[3]{#2}
\let\VANthebibliography\thebibliography
\def\thebibliography{\DeclareRobustCommand{\VAN}[3]{##3}\VANthebibliography}

\usepackage{graphicx}	
\usepackage{amsmath}	
\usepackage{color,soul} 

\usepackage[dvipsnames]{xcolor}

\title{Dark matter profiles of SPARC galaxies: a challenge to fuzzy dark matter}

\author[]{M. Khelashvili,$^{1,2,3}$\thanks{E-mail: khelashvili@fias.uni-frankfurt.de}  
A. Rudakovskyi,$^{2,4}$ 
S. Hossenfelder$^{1}$ 
\\ 
$^{1}$ Frankfurt Institute for Advanced Studies, Ruth-Moufang-Str. 1, Frankfurt am Main, 60438, Germany \\
$^{2}$ Bogolyubov Institute for Theoretical Physics of the NAS of Ukraine, Metrolohichna Str. 14-b, Kyiv, 03143, Ukraine \\
$^{3}$ Goethe Universit\"{a}t, Max-von-Laue Str. 1, Frankfurt am Main, 60438, Germany \\
$^{4}$Kyiv Academic University, 36 Vernadsky blvd., Kyiv, 03142, Ukraine\\
}

\date{Accepted XXX. Received YYY; in original form ZZZ}

\pubyear{2021}

\begin{document}
\label{firstpage}
\pagerange{\pageref{firstpage}--\pageref{lastpage}}
\maketitle

\begin{abstract}
Stellar and gas kinematics of galaxies are a sensitive probe of the dark matter distribution in the halo. The popular fuzzy dark matter models predict the peculiar shape of density distribution in galaxies: specific dense core with sharp transition to the halo. Moreover, fuzzy dark matter predicts scaling relations between the dark matter particle mass and density parameters. In this work, we use a Bayesian framework and several dark matter halo models to analyse the stellar kinematics of galaxies using the \textit{Spitzer Photometry \& Accurate Rotation Curves} database. We then employ a Bayesian model comparison to select the best halo density model. We find that more than half of the galaxies prefer the fuzzy dark model against standard dark matter profiles (NFW, Burkert, and cored NFW). While this seems like a success for fuzzy dark matter, we also find that there is no single value for the particle mass that provides a good fit for all galaxies.  
\end{abstract}




\section{Introduction}

Dark matter (DM) was discovered more than 80 years ago, but its nature has remained a mystery. It is usually assumed to be composed of unknown particles which are not contained in the Standard Model of particle physics. However, despite considerable efforts, the origin and the parameters of those hypothetical particles are still unknown. The possible mass of proposed dark matter particles depends strongly on the specific model and the range spans dozens of orders of magnitude.

The most widely used model of dark matter is cold dark matter ({\sc CDM}). It assumes a non-relativistic in the Early Universe, collisionless dark matter particles. 
This model has unquestionably been successful. Some of its achievements are predicting properties of the the large-scale structure and some aspects of {\sc CMB} anisotropies. {\sc CDM} is typically assumed to be a type of weakly interacting massive particle ({\sc WIMP}) with a mass of the order of GeV or larger. 

The focus of this paper is an alternative to cold dark matter, commonly called fuzzy dark matter ({\sc FDM}). These are axion-like particles which can appear for example in string theory  \cite[see, e.g.,][]{Arvanitaki2010}. 

Fuzzy dark matter is assumed to be composed of bosonic particles with masses $\sim10^{-24}\text{--}10^{-19}$\,eV and is the lightest dark matter candidate \citep[see, e.g.][]{Hu:00, ferreira2021ultralight, Hui2021}. {\sc FDM} particles with masses $\lesssim10^{-24}$ are ruled out by {\sc CMB} observations \citep{Hlozek:15}, while the small-scale structures abundance in FDM with mass $\gtrsim10^{-19}$\,eV are indistinguishable from CDM \citep{Hu:00}. The observed abundance of structures on small spatial scales is another way to constrain the {\sc FDM} parameters. This approach uses the Lyman-$\alpha$ forest \citep{Irsic:17, Nori:19, Rogers:2021}, the high-redshift galaxy luminosity functions \citep{Bozek:15,Schive:16, Corasaniti:17, Ni:19, Schutz:20}, and the Milky Way subhalos \citep{Nadler:21, Banik:21}.

{\sc FDM} possesses all the virtues of cold dark matter on cosmological scales but may have advantages on galactic scales.
This is because a system composed of ultra-light bosonic particles has a non-negligible quantum pressure arising from the uncertainty principle. This keeps the central density of halos finite and avoids the core-cusp problem. 
It also decreases the amount of low-mass structures and thereby suppresses the power spectrum on small scales \cite[see, e.g.,][]{Hu:00}, solving the missing satellite problem. \citep[see, e.g., extensive reviews][]{Weinberg:15, DelPopolo:17, Bullock:17}.

A prominent feature of a FDM halo is its peculiar density profile: a dense central soliton core with a sharp transition to the less dense outer region with granular structure \citep[see, e.g.,][]{Schive2014, Schive2014a}.  This shape significantly differs from other cored density profiles, e.g., CDM halos modified by the baryonic feedback \citep[see, e.g.,][]{Read2016a, Read2016b}, profiles of fermionic DM \citep[see, e.g.,][]{Shao:12, Maccio:12b, Savchenko2019} or density profiles of self-interacting DM \cite[see, e.g.,][]{Tulin2018}. Moreover, the {\sc FDM} model implies a scaling relation which links the central density, the characteristic radius of the central soliton, and the {\sc FDM} particle mass \citep[see, e.g.,][]{Schive2014}. Simulations also predict a relation between the virial halo mass and the soliton mass \citep[see, e.g.,][]{Schive2014a, Hui2017}.   The predicted granular structure of the halo also allows one to constrain the ultra-light dark matter parameters \citep{Amorisco:18, Dalal:21}. All this makes the observed kinematics of galaxies an appealing probe for the {\sc FDM} model.

In this paper, we analyse the rotation curves of the {\sc SPARC} galaxies \citep{Lelli2016} within a Bayesian inference framework \citep[see a review of Bayesian methods in cosmology in][]{Trotta:17}. With the {\sc SPARC} database we can test the predictions of various DM models for galaxies in a wide range of masses, luminosities, and morphological types. We use the Bayesian evidence to compare the {\sc FDM} halo model with the Navarro--Frenk--White ({\sc NFW}), cored {\sc NFW} (coreNFW), and Burkert density profiles. This statistical approach naturally includes a penalty for models with additional parameters and allows one to reject models that are unnecessarily complex. We also obtain the credible intervals of the fuzzy dark matter mass for different galaxies in the {\sc SPARC} sample and compare them with each other.

This paper is organised as follows: in Sec.~\ref{sec:method} we describe the dark matter profiles under consideration, the rotation curve model and the data analysis framework. In Sec.~\ref{sec:results} we report our findings and compare them to previously published results in Sec.~\ref{sec:prior} In Sec.~\ref{sec:discussion} we discuss our results, and we  conclude in Sec.~\ref{sec:conclusions}. Throughout this paper we use the Hubble parameter $H_0=73\,\frac{\text{km/s}}{\text{Mpc}}$.

\section{Methodology}\label{sec:method}

\subsection{Dark matter halo models}

\subsubsection*{NFW density profile}

CDM halos are well described by the Navarro--Frenk--White profile inferred from the DM-only $N$-body simulations \citep{Navarro:95, Navarro:96}:
\begin{equation}
    \rho_\text{NFW}(r) = \frac{\rho_n}{\frac{r}{r_n}\left(1+\frac{r}{r_n}\right)^2} \,,
    \label{eq:nfw-profile}
\end{equation}
where $r_n$ is the half-mass radius, and $\rho_n = 4 \rho_\text{NFW} (r_n)$ is a characteristic density.

\subsubsection*{coreNFW density profile}

Stellar feedback can affect the inner slope of the DM density profile in galaxies. In this way, a cusped halo may be smoothed to one with a core in the inner region.
Realistic high-resolution hydrodynamic simulations with baryonic feedback \citep{Read2016a, Read2016b}
show that the DM mass distribution in dwarf galaxies can be described by the following form:
\begin{equation}
    M_\text{coreNFW}(r) = M_\text{NFW}(r)\left(\tanh\frac{r}{r_c}\right)^n\,,
    \label{eq:core-nfw-profile}
\end{equation}
in which the mass $M_\text{NFW} (r)$ is calculated according to the initial NFW density profile. The effect of stellar feedback is fully incorporated here through the last factor, in which the parameter $n$ ranges from $0$ (corresponding to the NFW profile) to $1$ (describing a completely cored profile). The parameter $r_c$ in this expression represents the characteristic radius of the stellar component, which is proportional to the stellar half-mass radius. Both $n$ and $r_c$ are considered as free parameters in our analysis.

\subsubsection*{Burkert density profile}
We also consider the popular empirical Burkert density profile \cite{Burkert_1995}:
\begin{equation}
    \rho_\text{Burk}(r) = \frac{\rho_b}{\left(1+\frac{r}{r_b}\right)\left(1+\frac{r^2}{r_b^2}\right)}\,,
    \label{eq:burkert-profile}
\end{equation} 
where $r_b$ is the radius that contains $1/4$ of the mass and $\rho_b = \rho_\text{Burk}(0)$ is the density at the halo centre.  

\subsubsection*{Fuzzy dark matter density profile}\label{subsec:fdm}

A {\sc FDM} halo is composed of a dense central soliton surrounded by an envelope of incoherent phase with granular structure, according to the hydrodynamic dark matter only simulations \citep{Schive2014, Schive2014a, Veltmaat2016} and more sophisticated simulations such as finite-difference solving of the Schr\"{o}dinger--Poisson system of equations for a self-gravitating free scalar field \citep{mina2020solitons, schwabe2021deep}.  

The density distribution in the central soliton is described by the `boson-star' ground state solution of the Schr\"{o}dinger--Poisson 
system of equations \cite{Schive2014, Schive2014a}: 
\begin{equation}
    \rho_\text{sol}(r) = \frac{\rho_s}{\left(1 + 9.1\cdot 10^{-2} \left(\frac{r}{r_s}\right)^2\right)^8} \,,
    \label{eq:soliton}
\end{equation}
where $\rho_s$ is the density at the halo center, and $r_s$ is the characteristic scaling radius of the soliton, which contains approximately a quarter of the total soliton mass $M (r \le r_s) \approx \frac{1}{4} M_\text{sol}$. 

The {\sc FDM} model predicts a relation between $r_s$ and $\rho_s$ which arises from the exact internal scaling symmetry of the Schr\"{o}dinger--Poisson system~\citep[see, e.g.][]{Schive2014, mina2020solitons}.
It links the central density of the soliton to the characteristic radius and FDM particle mass $m = m_{22}\cdot 10^{-22}\text{eV}$ as follows: 
\begin{equation}
     \rho_s = \frac{1.9\cdot10^{-2}}{m_{22}^2 \left(\frac{r_s}{\text{kpc}}\right)^{4}} \frac{M_{\odot}}{\text{pc}^3}\,.  
    \label{eq:scaling}
\end{equation}
Note that according to our definition $m_{22}$ is dimensionless.

In the outer halo region, the structure is granular, and the 
density profile of the outer part is well described by the {\sc NFW} profile on average. Therefore, we will here use the following form of {\sc FDM} density profile: 
\begin{equation}
\rho_\text{FDM}(r) = 
    \begin{cases}
        \rho_\text{sol}(r), & r < r_a \, , \\
        \rho_\text{NFW}(r), & r > r_a \, ,
    \end{cases}
    \label{eq:fdm-profile}
\end{equation}
where $r_a$ is the transition radius at which the density profile changes its behaviour.
It was suggested in \citep{Schive2014, mina2020solitons} that $r_a = \alpha r_s$ with $\alpha\approx3$. However, this relation is not exact and may vary from halo to halo.

We also consider the following relation between the soliton and halo masses  \citep{Schive2014a, Hui2017}: 
\begin{equation}
     M_\text{sol} = 2.7 \cdot 10^8 \frac{\delta }{m_{22}} \left( \frac{M_\text{vir}}{10^{10} M_{\odot}} \right)^{1/3} M_{\odot}\,. 
    \label{eq:soliton-halo-scaling}
\end{equation}
It arises from the connection of the energy per unit mass of the central soliton and outer halo \citep[see, e.g.,][]{Bar_2018}. The factor $\delta$ here describes the uncertainty up to $\pm 50\%$ in this relation \citep{Schive2014a, Chan:22}.

\subsubsection*{Parametrization of the density profiles}

While the density $\rho_x$ and radius $r_x$ is a natural parametrization of the {\sc NFW} and Burkert density profiles, these parameters can vary in very wide ranges (orders of magnitude) from galaxy to galaxy. Because of this, following \cite{Li2020}, we use instead the rotation velocity $v_{200}$ and concentration $c_{200}$ as independent parameters. 
The velocity $v_{200}$ is defined as
\begin{equation}
v_{200}^2 = \frac{G_\text{N}}{r_{200}}\int^{r_{200}}_{0}{\frac{4\pi}{3}\rho_\text{DM} r^2\mathrm{d} r} \, ,
\end{equation}
where $r_{200}$ is the radius enclosing the region with average density $200$ times greater than the critical density of the Universe.

The concentration is defined as
\begin{equation}
c_{200}=\frac{r_{200}}{r_x}\,,
\end{equation}
where $r_x$ is the characteristic radius scale for the Burkert, {\sc NFW} and coreNFW profiles. 

Compared to the {\sc NFW} profile, the coreNFW profile includes two additional parameters $n$ and $\beta_c=r_c/r_n$ (see Eq.~\ref{eq:core-nfw-profile}), where $r_n$ is the characteristic radius of the initial {\sc NFW} profile.\footnote{Note that we consider $n$ and $\beta_c$ as free parameters, in contrast with \cite{Li2020}.}

In general, the {\sc FDM} profile can be described by four parameters: $\rho_s$ and $r_s$ for the central soliton, the transition radius $r_a$, and $r_n$  for the {\sc NFW} `tail' (the {\sc NFW} parameter $\rho_n$ is fixed by the continuity condition).
For convenience, we introduce the following baseline parametrization: logarithm of the mass of the FDM particle $\log_{10} m_{22}$, the ratio between the transition and soliton radii $\alpha = r_a/r_s$, the parameter $\delta$ (see Eq.~\ref{eq:soliton-halo-scaling}) and the rotation velocity $v_{200}$. The parameters $v_{200}$, $\log_{10} m_{22}$, $\alpha$, $\delta$ can be directly translated to $\rho_s$, $r_s$, $\rho_n$, $r_n$ via the relations in Eq.~\ref{eq:scaling}, \ref{eq:soliton-halo-scaling}; 
see more in Appendix~\ref{appendix:FDM-parametrization}.

We list the free parameters of each DM model in Tab.~\ref{tab:model_parameters}.

\begin{table}
    \centering
    \begin{tabular}{l|l}
        \hline
         Model & Parameters \\
        \hline
         NFW & $v_{200}\,, \quad  c_{200}$ \\
         Burkert & $v_{200}\,, \quad c_{200}$ \\
         coreNFW & $v_{200}\,, \quad c_{200}\,, \quad n\,, \quad \beta_c$ \\
         FDM &  $v_{200}\,, \quad \log_{10} m_{22}\,, \quad \alpha\,, \quad \delta$\\
    \end{tabular}
    \caption{Free parameters of the DM models under consideration.}
    \label{tab:model_parameters}
\end{table}

\subsection{Rotational velocity model}

We use the model of \cite{Li2020} to calculate the rotational velocity. The total gravitational force is the sum of the attraction forces of luminous and dark matter; therefore, the observed rotation velocity is 
\begin{equation}
v = \sqrt{v_\text{lum}^2 + v_\text{DM}^2} \,,
\end{equation}
where $v_\text{DM}^2$ and $v_\text{lum}^2$ are the contributions from dark and luminous matter, respectively. 

The luminous matter contribution is defined by 
\begin{equation}
v^2_\text{lum} = \left|v_\text{gas}\right|v_\text{gas} + \Upsilon_\text{disk} \left|v_\text{disk}\right|v_\text{disk} + \Upsilon_\text{bulge}\left|v_\text{bulge}\right|v_\text{bulge} \,,
\label{eq:mass_model}
\end{equation}
where $\Upsilon_\text{disk}$ and $\Upsilon_\text{bulge}$ are mass-to-light ratios of the disk and bulge, respectively.\footnote{In expression Eq.\,\ref{eq:mass_model}, some baryonic components may contribute with negative sign to the total acceleration. This may be caused by the strong depression of the gas distribution in the innermost regions of some galaxies. In this case, the gravitation from the outer regions leads  to the negative contribution to the total acceleration. To take this into account, the SPARC database sometimes provides negative velocity contribution from the baryonic component in the inner regions of galaxies.}

\subsection{Data analysis}

We analyse the rotation curves of galaxies from the Spitzer Photometry \& Accurate Rotation Curves (SPARC) database \citep{Lelli2016}. It provides detailed 3.6\,$\mu$m photometry and rotational velocity data for 175 galaxies. The SPARC database includes galaxies of different Hubble types, spanning wide range of masses and luminosities. This allows one to robustly test the scaling relations predicted by the model of FDM halos.   

Our model includes the DM density profile parameters (up to four parameters), the mass-to-light ratios $\Upsilon_\text{disc}$ and $\Upsilon_\text{bulge}$, the inclination angle $i$, and the distance to the galaxy $D$.

The observed total velocity and its uncertainty depends on the galaxy inclination angle $i$ as \citep{Li2018}
\begin{equation}
v_\text{obs} \propto 1 / \sin i, \quad \delta v_\text{obs} \propto 1 / \sin i \,.
\end{equation}
The contribution of the luminous matter to the rotational velocity depends on the distance to the galaxy $D$ \citep{Li2018} as
\begin{equation}
v_\text{disk, bulge, gas} \propto \sqrt{D} \,.
\end{equation}
The radius also scales with the distance:
\begin{equation}
r \propto D\,.
\end{equation}

Since our velocity models have up to eight parameters, we select galaxies with at least eight data points. We also follow \cite{Lelli2016} and do not include galaxies with poor `quality' of the data.

We specifically focus on low surface brightness (LSB) galaxies. They are useful for constraining the dark matter properties because they appear to be dark matter dominated, even at small radii. We take into account that LSB galaxies have surface brightness $ \log B_\text{eff} \le 1.5 L_{\odot} / \text{pc}^2 $ (Pengfei Li, private communication).   

We apply the Bayesian inference for constraining the model parameters and for model selection. According to the Bayes theorem, the posterior probability density of the model parameters $\theta$ is given by
\begin{equation}
    P(\theta|d, \mathcal{M}) = \frac{P(d|\theta, \mathcal{M})\,\pi(\theta, \mathcal{M})}{p(d|\mathcal{M})} \, ,
\end{equation}
where $P(d|\theta, \mathcal{M})$ is the likelihood of the data $d$, $P(d|\mathcal{M})$ is the marginalized likelihood (evidence), and $\pi(\theta, \mathcal{M})$ is the prior probability density. 

The evidence is defined as
\begin{equation}
    Z \equiv P(d|\mathcal{M}) = \int P(d|\theta, \mathcal{M})\pi(\theta ) \mathrm{d}\theta \, .
\end{equation}
We assume the likelihood probability density in the Gaussian form
\begin{equation}
P(D|\theta, \mathcal{M})=\prod_\text{i} \frac{1}{\sqrt{2\pi}\sigma_\text{i}}\mathrm{exp}\left(-\frac{\left( v_\text{i,obs}-v_\text{pred}(\theta, \mathcal{M})(r_\text{i}) \right)^2}{2\sigma_\text{i}^2}\right)\,,
\end{equation}
where $v_\text{i,obs}$ is the observed rotational velocity at radius $r_\text{i}$, $\sigma_\text{i}$ is the error of its determination, and $v_\text{pred}(\theta, \mathcal{M})(r)$ is the rotational velocity predicted by model $\mathcal{M}$ with parameters $\theta$. 

We adopt the prior choice made in \cite{Li2020} for $c_{200}$ and $v_{200}$. We consider maximally wide uniform priors for the FDM parameters $\alpha$ and $\delta$ to capture the results of the FDM simulations \citep{Schive2014, Schive2014a, mina2020solitons, Nori:20, Chan:22}. The lower bound of $-3$ in the prior for $\log_{10} m_{22}$ is motivated by the CMB constraints \citep{Hlozek:15}, while FDM with  $\log_{10} m_{22}=3$  is similar to CDM.
We summarize the information about the parameters of the models and their priors in Tab.~\ref{tab:priors}. The functional forms of the priors are described in Appendix~\ref{sec:priors}.

\begin{table*}
    \centering
    \begin{tabular}{cccccc}
\hline
Parameter & Fiducial value & Std dev & Units & Allowed range & Prior \\
\hline
\multicolumn{6}{c}{Dark matter} \\
\hline
$v_{200}$ & -- & -- & km / s &  10 -- 500  & uniform \\
$c_{200}$ & -- & -- & -- & 0 -- 1000 & uniform\\
$n$ & -- & -- & -- & 0 -- 1 & uniform \\
$\beta_c$ & -- & -- & kpc & 0 -- 1 & uniform \\
$\log_{10} m_{22}$ & -- & -- & -- & $-3$ -- 3& uniform \\
$\alpha$ & 3 & -- & -- & 1 -- 7 & uniform \\
$\delta$ & 1 & -- & -- & 0.5 -- 1.5 & uniform \\
\hline
\multicolumn{6}{c}{Galaxy astrophysics} \\
\hline
$\Upsilon_\text{disk}$ & 0.5 & 0.1 dex &$M_{\odot}/L_{\odot}$ & -- & log-normal \\
$\Upsilon_\text{bulge}$ & 0.7 & 0.1 dex  &$M_{\odot}/L_{\odot}$ & -- & log-normal \\
$D$ & SPARC & error & Mpc & -- & normal \\
$i$ & SPARC & error & deg & -- & normal \\
\hline
\end{tabular}
    \caption{The parameters of the models and the corresponding priors. The free parameters of each DM model are listed in Tab.~\ref{tab:model_parameters}. We assume the same range of similar parameters in different models. }
    \label{tab:priors}
\end{table*}

We use the dynamic nested sampling \citep{Higson:19} implemented in \textsc{dynesty} \textsc{Python} package \citep{dynesty} for the estimation of evidences and posteriors.

Knowing the evidences allows us to compare models $\mathcal{M}_1$ and $\mathcal{M}_2$ via the Bayes factor:
\begin{equation}
    K_{12} = \frac{P(\mathcal{M}_1|D)}{P(\mathcal{M}_2|D)} = \frac{P(D|\mathcal{M}_1)P(\mathcal{M}_1)}{P(D|\mathcal{M}_2)P(\mathcal{M}_2)} = \frac{P(D|\mathcal{M}_1)}{P(D|\mathcal{M}_2)}.
    \label{bayes_factor}
\end{equation}
If $\log_{10} K_{10}$ is between $1/2$ and $1$, then model $\mathcal{M}_1$ is regarded to be substantially supported by the data against $\mathcal{M}_2$;  if $\log_{10} K_{12}>1$, then $\mathcal{M}_1$ is regarded to be strongly supported \citep{Jeffreys:1939}.

\section{Results}\label{sec:results}

\subsection{Model comparison}
\begin{figure*}
    \centering
    \includegraphics[width = 1.6\columnwidth]{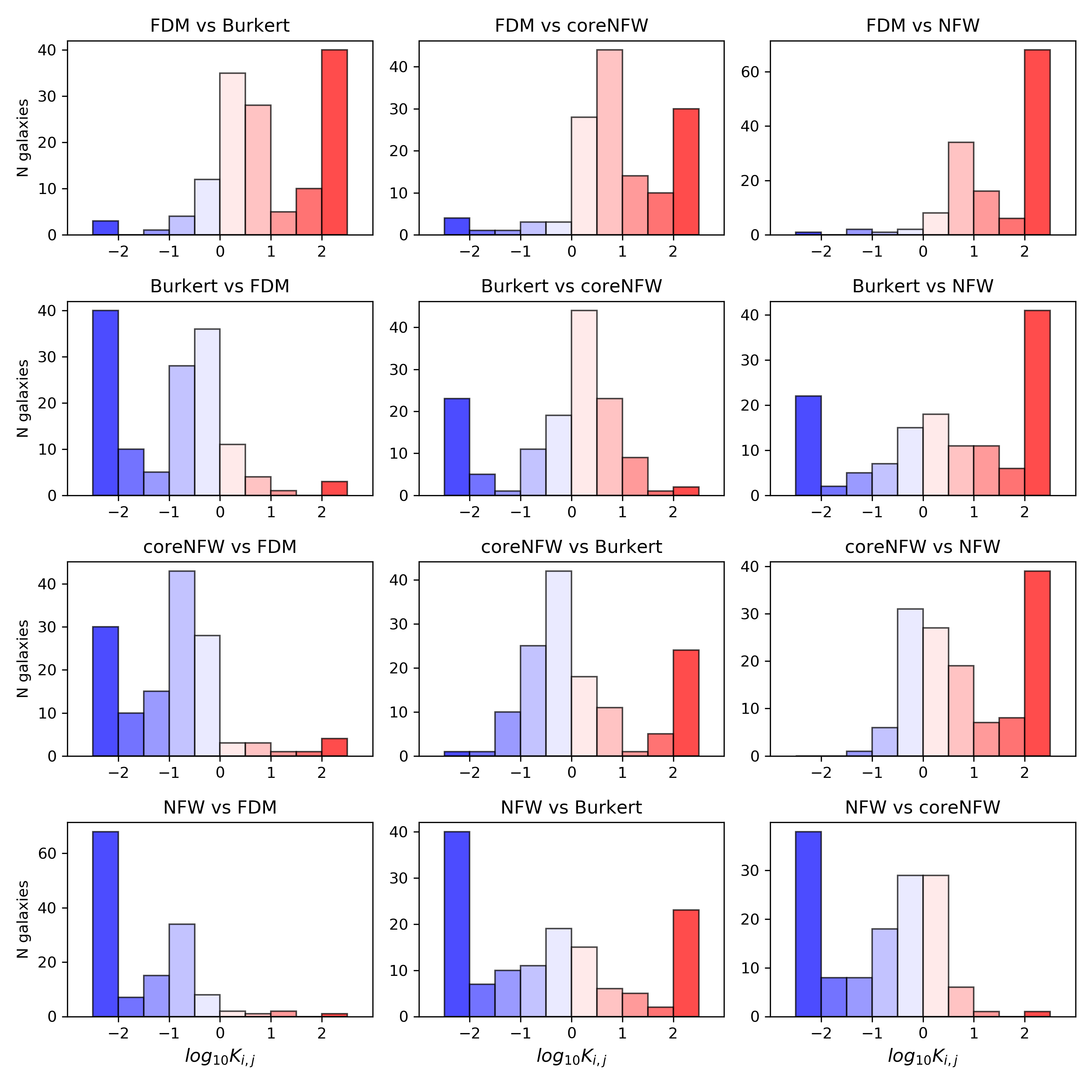}
    \caption{The number of galaxies that prefer the first model (red columns) or disfavour it (blue columns) as compared to the second model and the strength of preference.}
    \label{fig:pref_tab}
\end{figure*}

We first investigate which of the DM halo models fits the observed kinematics better. For this, we calculate the marginalised likelihoods for every galaxy and for each DM halo model under consideration. As mentioned above, this allows one to determine models that describe the data better in general. After that, we compute the Bayes factors for each pair of models.  The distribution of the Bayesian coefficients for each pair of models is shown in Fig.~\ref{fig:pref_tab}. 
 
We summarise the number of galaxies that substantially prefer or disfavor each DM halo model compared to other ones in Tab.~\ref{tab:pref_models_ngalaxy}. It clearly shows that half of the galaxies (72 out of 143) prefer the {\sc FDM} profile compared to all other DM density profiles under consideration.

\begin{table}
    \centering
    \begin{tabular}{|c|c|c|c|}
    \hline
    & Prefer & Disfavour & Indifferent  \\
    \hline
    FDM & 72 (15) & 13 (5) & 53 (22) \\
    Burkert & 5 (2) & 84 (20) & 49 (20) \\
    coreNFW & 2 (1) & 104 (29) & 32 (12) \\
    NFW & 0 (0) & 128 (38) & 10 (4) \\
    \hline
    \end{tabular}
    \caption{The number of galaxies that strongly prefer or disfavour each dark matter model. The numbers in parentheses correspond to the subsample of LSB galaxies.} 
    \label{tab:pref_models_ngalaxy}
\end{table}

\subsection{Astrophysical properties of the galaxies preferring FDM}

\begin{table}
    \centering
    \begin{tabular}{l|c|c}
        \hline
        Quantity & $R$ & $p$-value \\
        \hline
        Flat velocity & 0.35 & $9\cdot 10^{-5}$ \\ 
        Total luminosity & 0.32 &  $1\cdot 10^{-4}$ \\
        Surface Brightness & 0.2 & $0.02$ \\
        \hline
    \end{tabular}
    \caption{Pearson correlation coefficient $R$  between the galaxy feature and the preference of the FDM density profile by this galaxy. }
    \label{tab:correlation_coefficients}
\end{table}

\begin{figure*}
    \centering
    \includegraphics[width = 2\columnwidth]{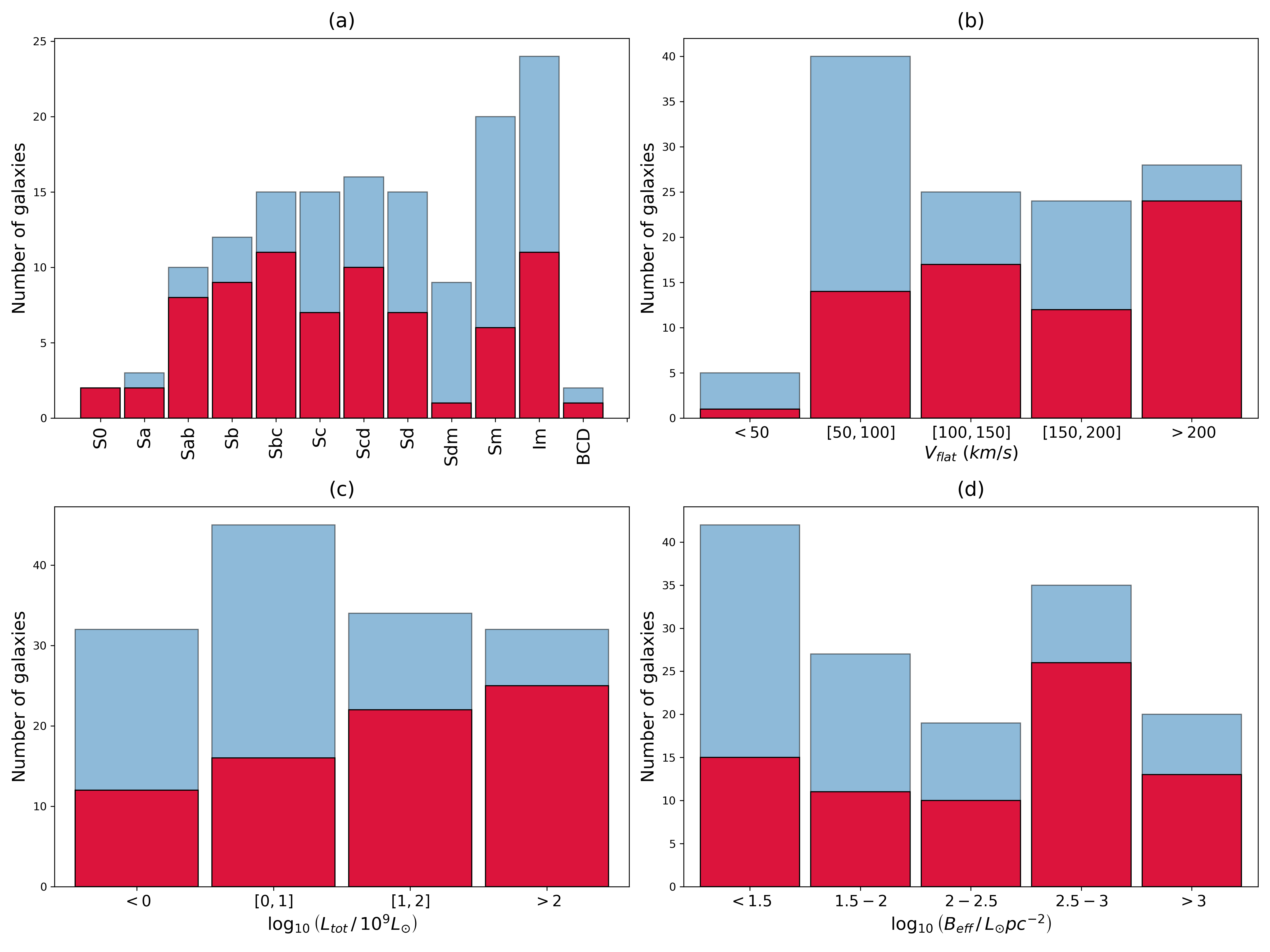}
    \caption{The fraction of the galaxies that prefer FDM among the galaxies of each Hubble morphological type (a), flat velocity (b), total luminosity(c) and effective brightness (d). In histogram (b) galaxies that do not reach flat velocity on the observed rotation curve are filtered out.} 
    \label{fig:FDM_preference_correlations}
\end{figure*}

Now we investigate correlations of the preference of the {\sc FDM} profile with the properties of galaxies. As general characteristics of galaxies, we choose the morphological Hubble type, the total luminosity, the effective surface brightness, and the flat velocity (the last quantity is introduced only for galaxies that 
satisfy the velocity flatness criteria of \cite{Lelli2015} at the external point of the rotation curve). Their values are provided by the {\sc SPARC} database and are model-independent. 

Correlations are illustrated by histograms in Fig.~\ref{fig:FDM_preference_correlations}. Red columns show the number of galaxies that substantially prefer the {\sc FDM} model, while blue columns correspond to the total number of galaxies.  We bin the continuous variables so that each bin has similar width while the morphology classification provides the natural binning. 

Our analysis reveals that early Hubble spirals and lenticular galaxies prefer the {\sc FDM} density profile among the profiles under investigations more often than late spirals or irregular galaxies. At the same time, among the continuous variables, the correlation is the strongest with the flat velocity, which correlates with the galaxy mass. 
Furthermore, we observe some correlation of the {\sc FDM} preference
with the total galactic luminosity, and a somewhat weaker correlation with the effective surface brightness. The latter correlation can easily be understood: massive galaxies often are more luminous and have higher surface brightness. \footnote{However, the more massive galaxy does not always have to be more luminous or to has greater surface brightness.}

For a better quantitative estimate, we calculate the Pearson correlation coefficient between the {\sc FDM} preference and each of the galactic parameters (flat velocity, total luminosity and effective surface brightness). We assign a discrete variable for the {\sc FDM} preference, which is chosen to be equal to $1$ if a galaxy strongly prefer {\sc FDM} and $0$ otherwise. The resulting correlation coefficients $R$
with $p$-values (the probabilities that the data points are uncorrelated) are given in Tab.\,\ref{tab:correlation_coefficients}.
The correlation coefficients and the corresponding $p$-values are calculated with \textsc{pearsonr} from \textsc{scipy.stats} \textsc{python} package.

These findings also confirm the presence of correlation between the astrophysical parameters of galaxies and their tendency to prefer {\sc FDM}. More luminous and more massive galaxies as well as galaxies of early Hubble types and those with greater surface brightness prefer the {\sc FDM} profile more often.  The small $p$-values may be interpreted as a sign that the correlations are not accidental. However, the correlations are quite weak (with the correlation coefficients around $0.3$).

\subsection{Fuzzy dark matter particle mass}%

The nested sampling algorithm also gives us the posterior probability distribution of the model parameters. We provide an example of the posterior distribution of the {\sc FDM} parameters for NGC~3741 in Appendix~\ref{sec:posterior}. As expected, we find strong degeneracy between the astrophysical and DM halo parameters.

\begin{figure*}
    \centering
    \includegraphics[width=2\columnwidth]{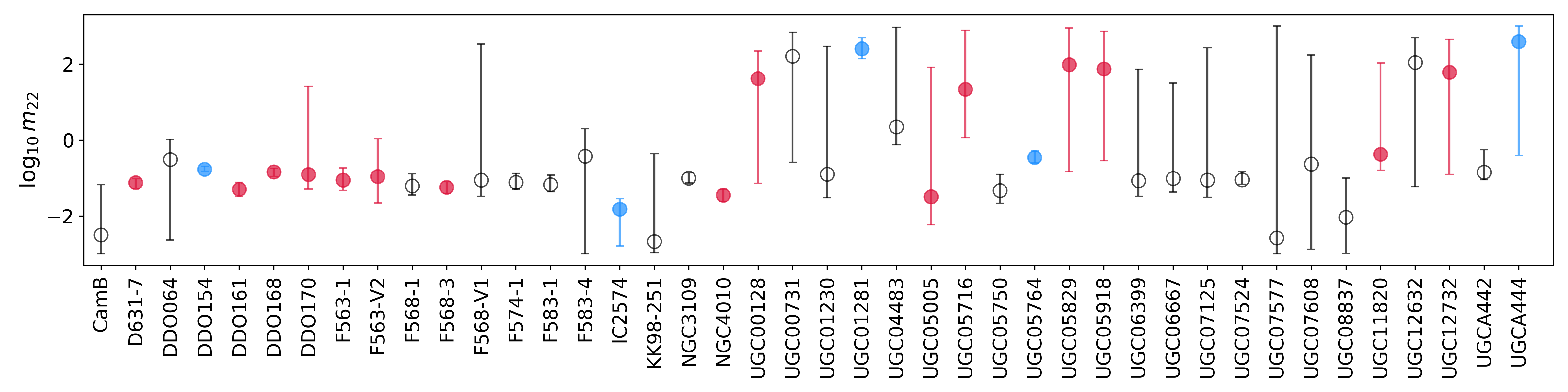}
    \caption{95\% credible intervals for the parameter $\log_{10} m_{22}$  for LSB galaxies from the SPARC catalogue (red colour indicates the galaxies that strongly prefer FDM, blue colour indicates those that disfavour FDM, and grey colour indicates the galaxies without strong preference).}
    \label{fig:m22_lsb}
\end{figure*}

We focus on the constraints on the mass of fuzzy dark matter particle arising from individual galaxies. Since {\sc LSB} galaxies are expected to be DM dominated, we use them for deriving the credible intervals for the mass parameter $\log_{10} m_{22}$ for individual objects. We define the credible interval as a single highest-density interval, see Appendix~\ref{appendix:hpd}. The obtained credible intervals are shown in Fig.~\ref{fig:m22_lsb}. 

One can clearly see that 95\% credible intervals for the mass parameter for individual galaxies are in tension with each other.

We find a similar tension for the $\mathrm{log}_{10}m_{22}$ credible intervals from individual galaxies with intermediate and high surface brightness, $ \log B_\text{eff} > 1.5 L_{\odot} / \text{pc}^2 $. 
 
\begin{figure}
    \centering
    \includegraphics[width = \columnwidth]{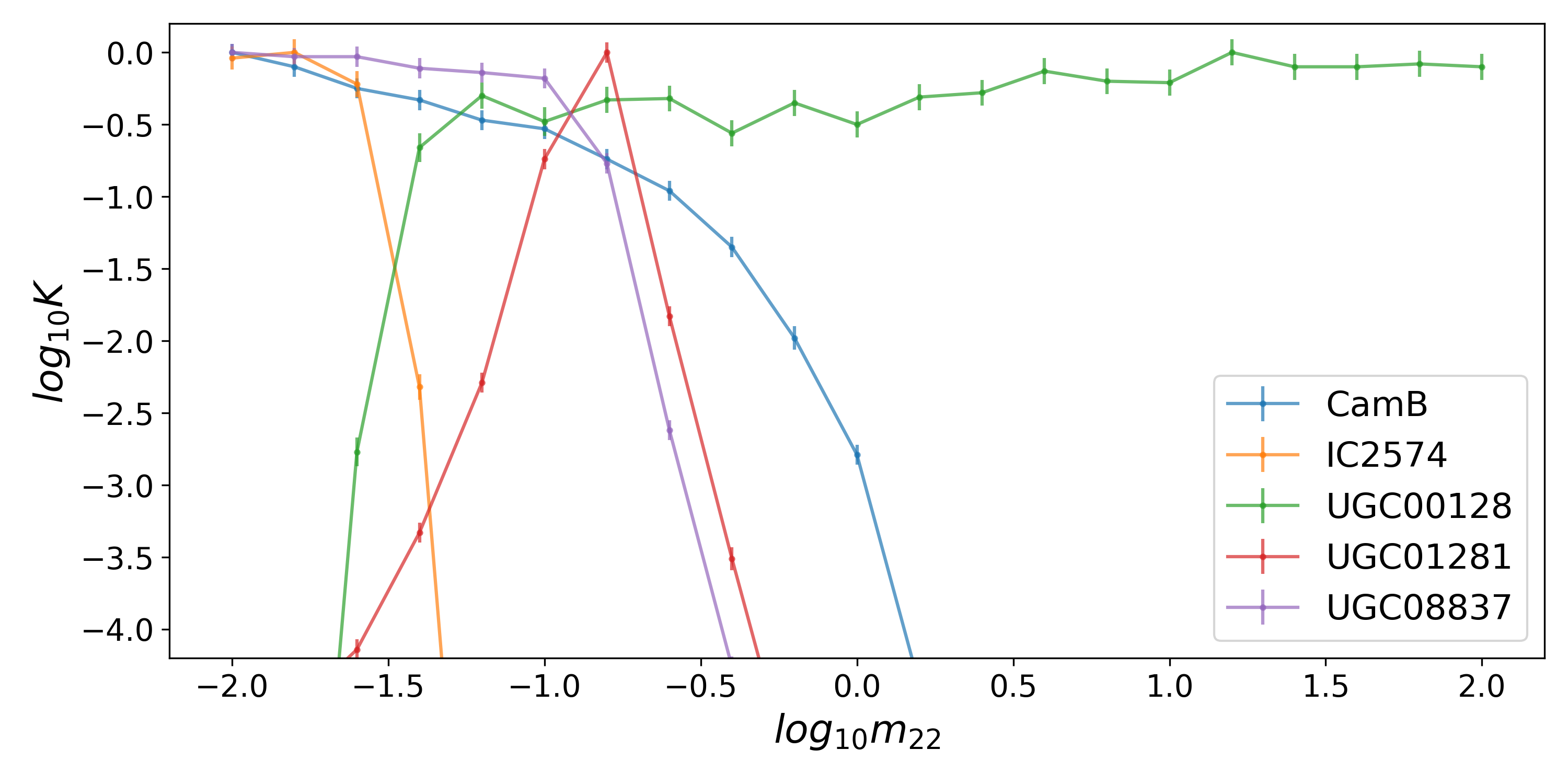}
    \caption{The logarithm of the Bayes factors for the FDM model with fixed values of the particle mass.}
    \label{fig:m22-bayes-vidence}
\end{figure}

Different definitions of credible interval may give different values. We therefore check the reliability of our results with the Bayesian model comparison. For this, we calculate the Bayesian evidences for the {\sc FDM} model with 21 fixed masses evenly spaced in the range $- 2 \le \log_{10} m_{22} \le 2$. For each galaxy, we select the {\sc FDM} model with the maximal evidence and compute the Bayes factors for the {\sc FDM} model with other masses against this `best' model. The logarithm of the Bayesian factors for CamB, IC2574, UGC00128, UGC01281, UGC08837 are shown in Fig.~\ref{fig:m22-bayes-vidence}. One can see that the FDM masses which are preferred for one particular galaxy are essentially disfavoured for some other galaxy in this collection ($\log_{10} K \leq -1.0$).

\subsection{Scaling relations of FDM} 

\begin{figure}
    \centering
    \includegraphics[width = \columnwidth]{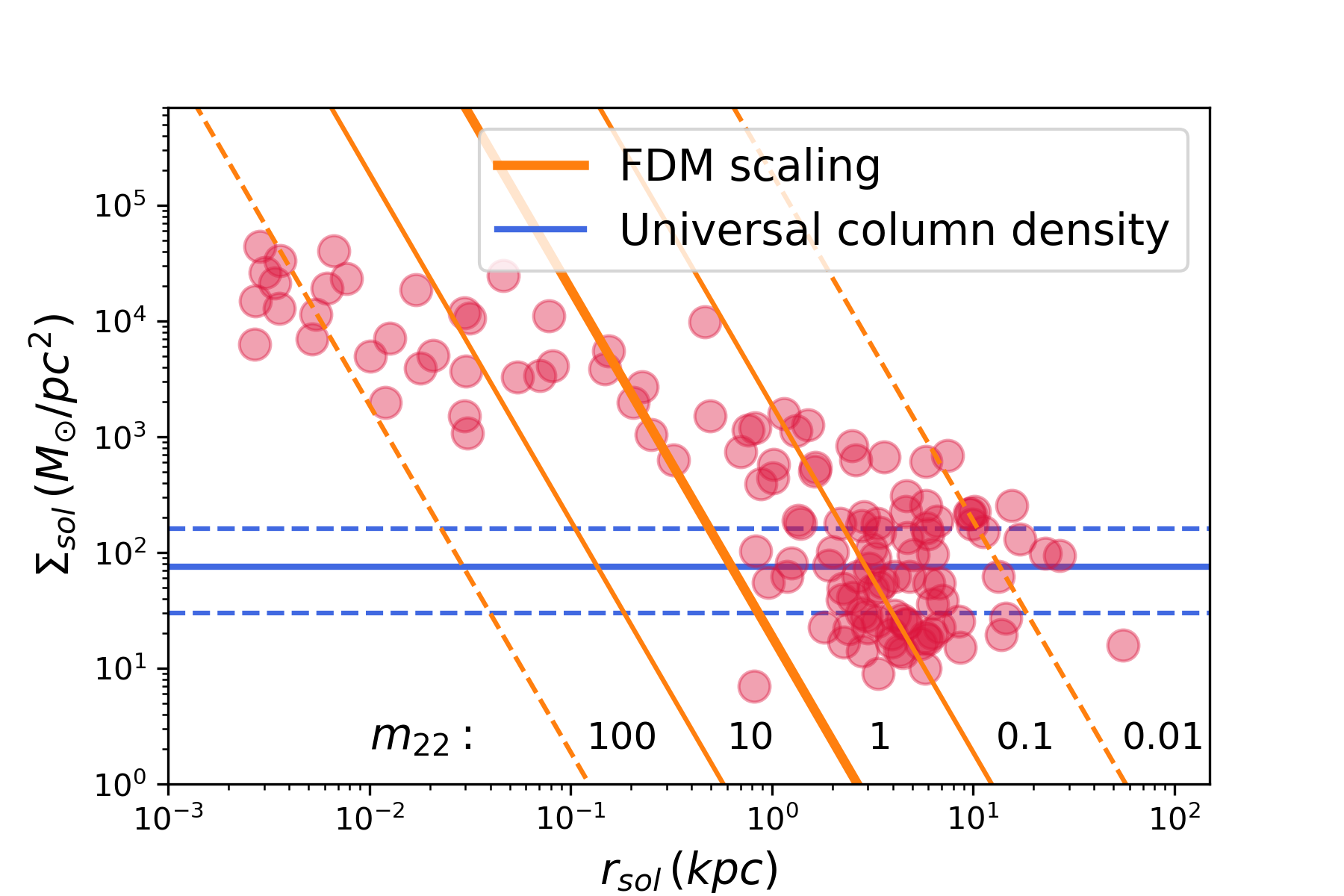}
    \caption{The core column density vs the characteristic soliton radius for galaxies from the SPARC database. The dots correspond to the MAP parameters found via Bayesian inference.  The predictions from the FDM scaling relation \eqref{eq:scaling} for particle masses in the range $ 10^{-24}\text{--}10^{-20} $~eV is shown by orange lines. }
    \label{fig:column_density}
\end{figure}

If FDM comprises all of the DM in the universe, the soliton density profile parameters should follow the scaling relation Eq.\,\ref{eq:scaling}. For an illustration of the comparison of our findings with the FDM prediction, we plot the best-fit core column densities $\Sigma = \rho_s \cdot r_s$  vs the core characteristic radii in Fig.~\ref{fig:column_density}. The red dots correspond to the maximum a posteriori (MAP) parameters for each galaxy. The yellow lines are the prediction of the FDM scaling relation Eq.\,\ref{eq:scaling} for different FDM masses $m_{22}$.  One observes a huge scatter in the values of the core column density relative to the theoretically predicted by Eq.\,\ref{eq:scaling}. Moreover, the best-fit value of $\Sigma_\text{sol}$ significantly differs from the universal core column density $\Sigma = 75^{+55}_{-45}M_\odot/\text{pc}^2$ (blue lines in Fig.~\ref{fig:column_density}) reported in \citep{Salucci:00, Donato:09, Kormendy:16}, in contradiction with the findings of \cite{Burkert:20}.\footnote{Note that the universal core column density was derived for the usual (e.g., Burkert or isotermal) DM profiles, which are quite different from the FDM profile.}

\begin{figure}
    \centering
    \includegraphics[width = \columnwidth]{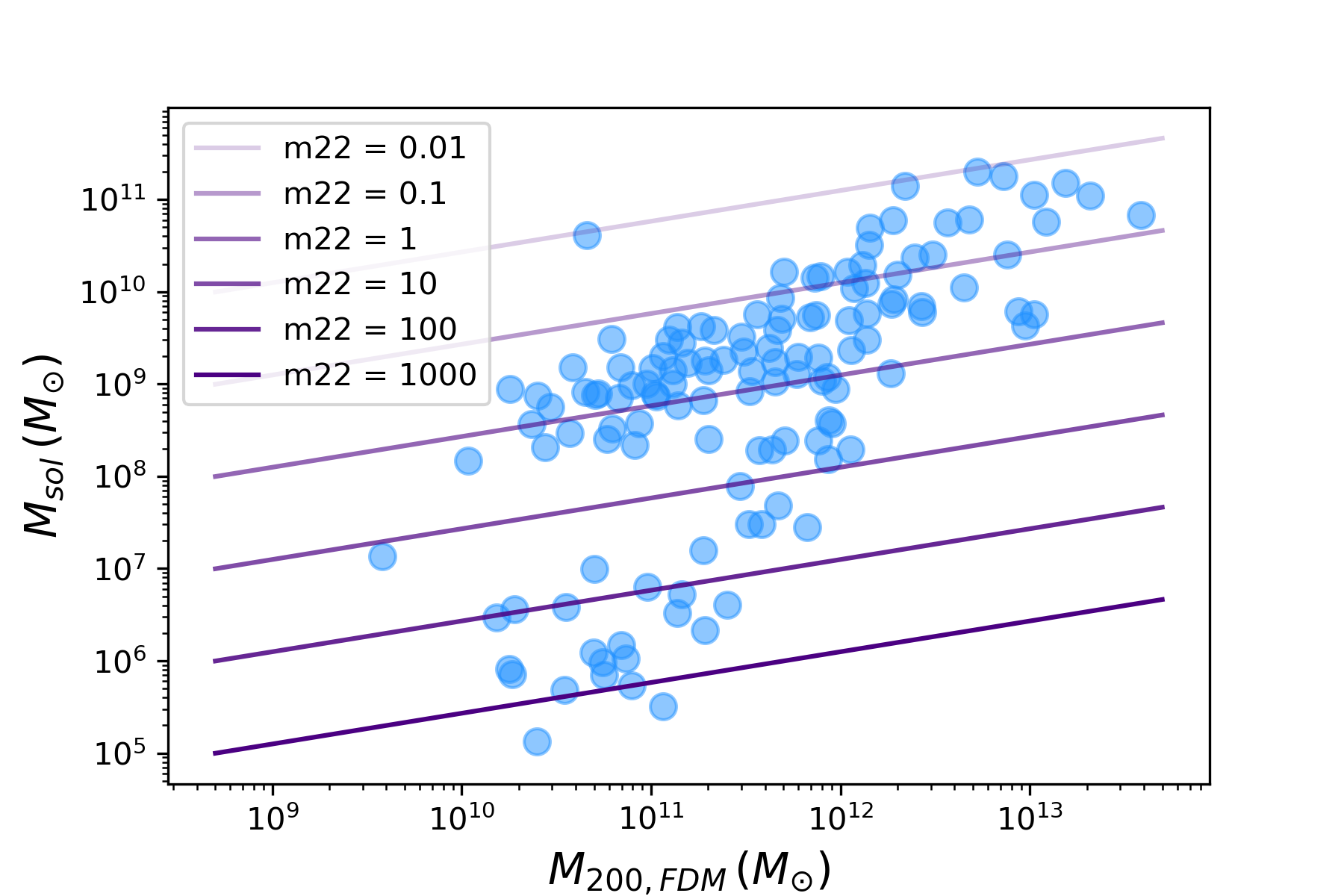}
    \caption{The mass of the FDM soliton against the virial halo mass for SPARC galaxies. The dots correspond to the MAP parameters found via Bayesian inference.  The solid lines are the FDM scaling relation \eqref{eq:soliton-halo-scaling} for particle masses in the range $ 10^{-24}\text{--}10^{-19} $~eV. }
    \label{fig:soliton-halo-mass}
\end{figure}

Our analysis also reveals a discrepancy between the theoretically predicted soliton--halo mass relation and the obtained MAP soliton mass $M_\text{sol}$ and virial mass $M_{200}$ (Fig.~\ref{fig:soliton-halo-mass}). The blue dots show the maximum a posteriori values obtained for different galaxies. The purple lines are based on Eq.\,\ref{eq:soliton-halo-scaling} with $\delta = 1$ from the simulations by \cite{Schive2014, Schive2014a}.

\begin{figure*}
    \centering
    \includegraphics[width = 2\columnwidth]{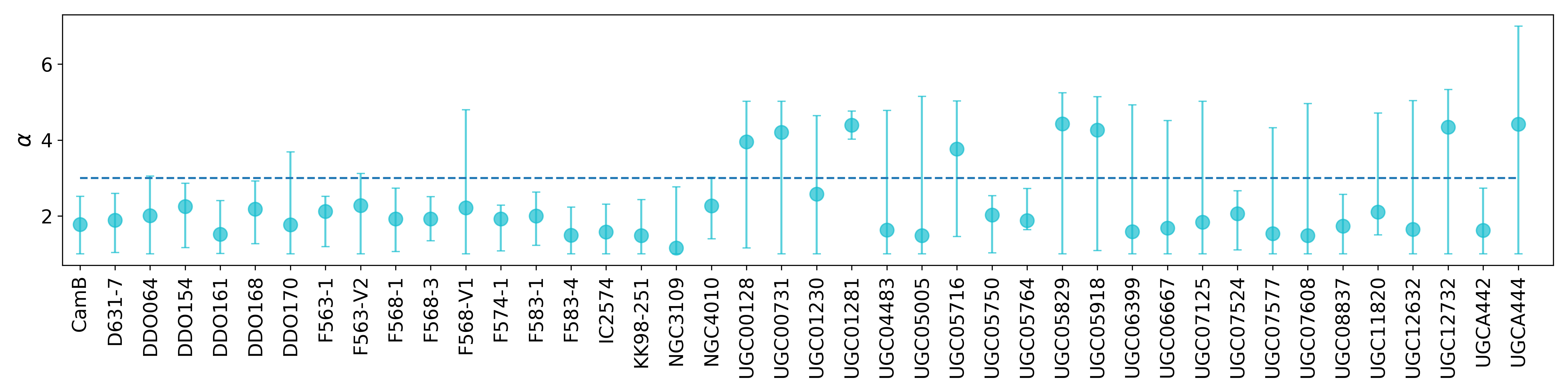}
    \caption{Credible $2 \sigma$ intervals for the transition parameter $\alpha$ of the FDM model for LSB galaxies from SPARC.}
    \label{fig:alpha_junction_LSB}
\end{figure*}

\begin{figure}
    \centering
    \includegraphics[width = \columnwidth]{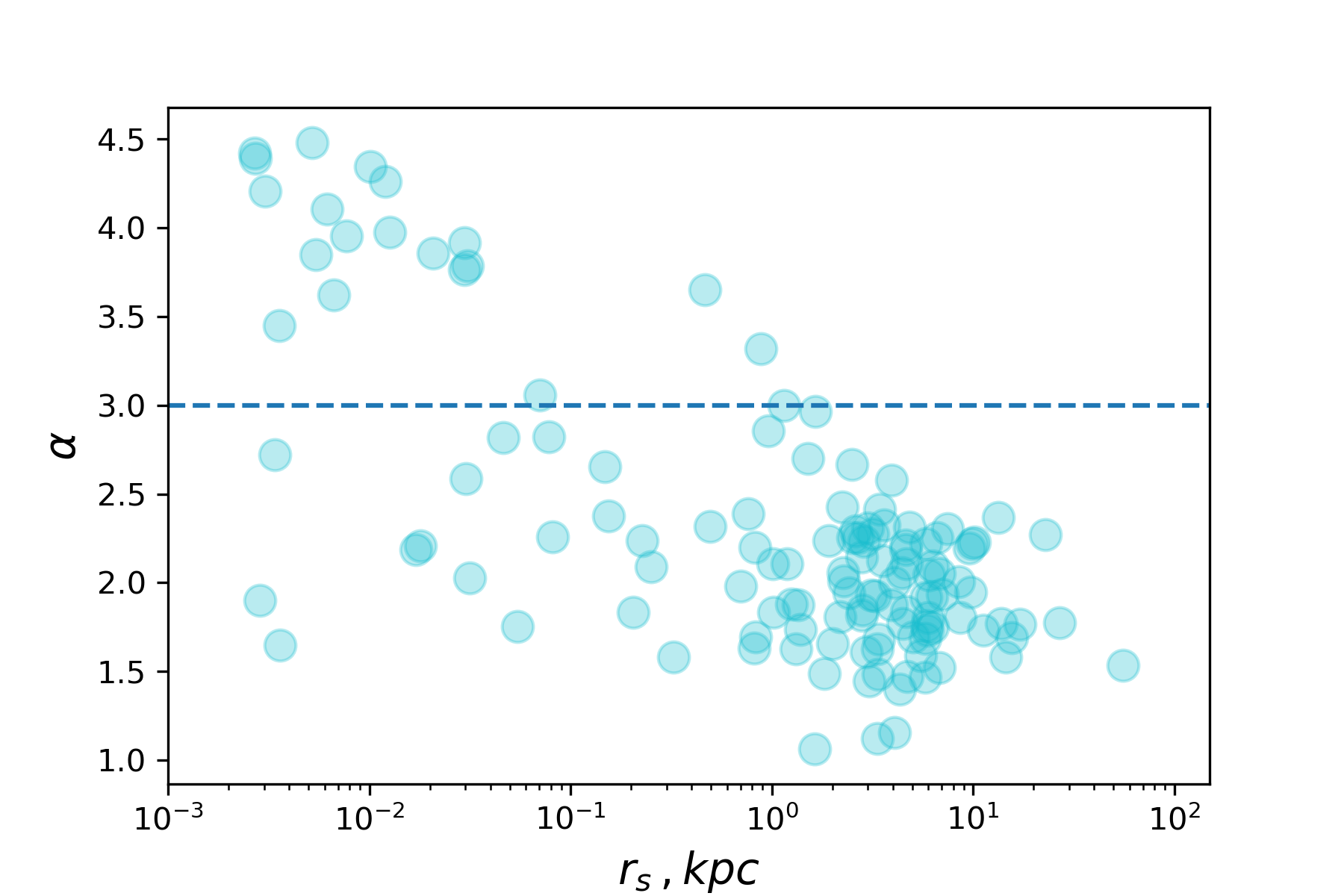}
    \caption{The transition radius parameter $\alpha$ vs the core radius for SPARC galaxies. The dots correspond to the MAP parameters found via Bayesian inference.}
    \label{fig:alpha_junction}
\end{figure}

We also find that the MAP values of $\alpha$, the ratio of the transition radius to the soliton radius, differ from the predictions of the simulations. We show the obtained credible intervals of $\alpha$ for LSB galaxies in Fig.~\ref{fig:alpha_junction_LSB}.  The best-fit values of $\alpha$ vs the core radius for all analysed galaxies are illustrated in Fig.~\ref{fig:alpha_junction}. Interestingly, lower values of $\alpha$ (less prominent central soliton) often correspond to more extended cores.

\subsection{The inner slope of the halo density profile}

\begin{figure*}
     \centering
     \includegraphics[width = 2\columnwidth]{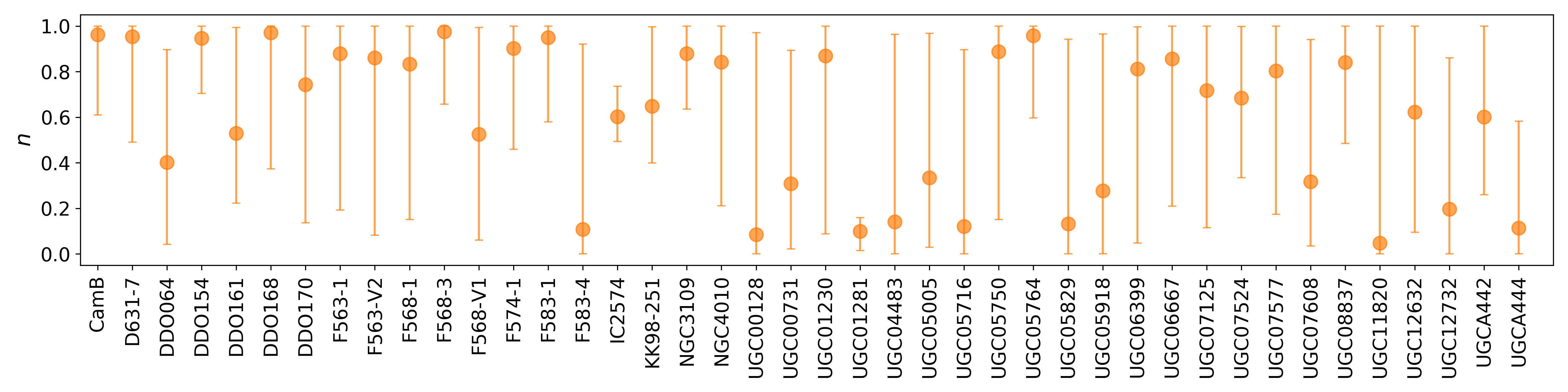}
     \caption{Credible $1\sigma$ region for ``degree of cuspiness'' parameter $n$ of coreNFW profile for LSB galaxies from SPARC database.}
     \label{fig:n_coreNFW}
 \end{figure*}

\begin{figure}
    \centering
    \includegraphics[width = \columnwidth]{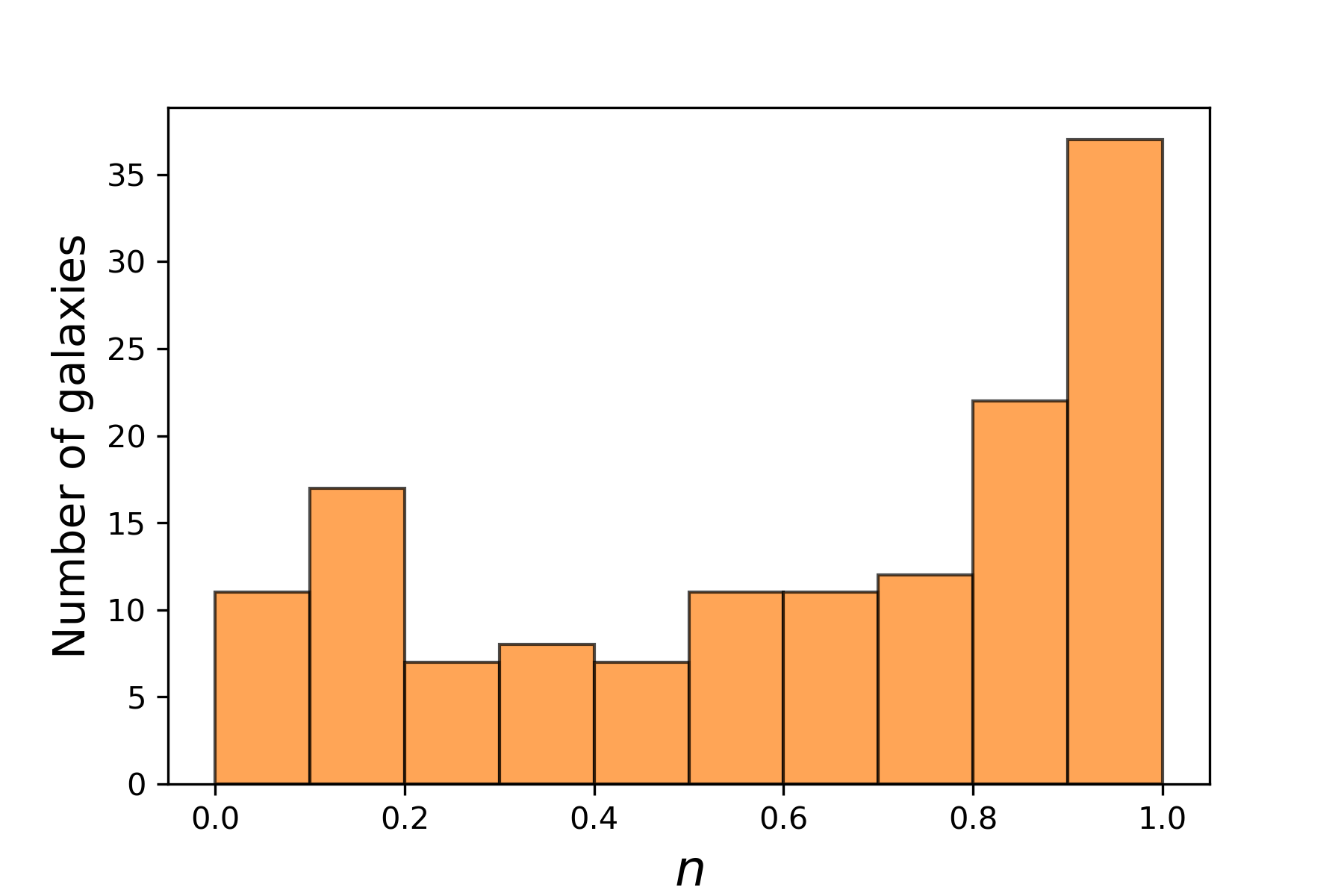}
    \caption{Histogram of 
    MAP value of $n$ coreNFW parameter for a sample of galaxies from SPARC database ($n$ close to $0$ corresponds to a cusp-like profile, most galaxies are better described by core-like profile $n \sim 1$).}
    \label{fig:n_coreNFW_hist}
\end{figure}

Are dark matter halo profiles cored or cusped? Our Bayesian model comparison does not reveal any galaxy with a strong preference for a cusped {\sc NFW} profile. Moreover, the {\sc NFW} density profile is disfavoured for most of the galaxies in the sample, 128 out of 138 (see Tab.~\ref{tab:pref_models_ngalaxy}). This result is in qualitative agreement with the findings of \cite{Li2020}.
The coreNFW profiles naturally describe DM halos with different `degree of cuspiness': the pure NFW profile corresponds to the parameter $n=0$, while $n=1$ corresponds to a completely cored density profile, see Eq.\,\ref{eq:core-nfw-profile}.  
We find that the credible intervals for the parameter $n$ are very large and often include values for both core-like and cusp-like halos. As an illustration, we provide 68\% credible intervals for $n$ 
in Fig.~\ref{fig:n_coreNFW}. Nevertheless, galaxies with the maximal a posteriori value of $n$ close to $1$ are much more frequent; see the histogram in Fig.~\ref{fig:n_coreNFW_hist}.

\section{Comparison to prior work}\label{sec:prior}

Our results are complementary to several constraints  on {\sc FDM} that were previously derived from the {\sc SPARC} data.

\cite{Bernal_2017} obtained a confidence interval of $0.0212\leq m_{22}\leq270$ for the mass of {\sc FDM} particles from the analysis of rotation curves of 18 {\sc LSB} and 6 (4 from SPARC database) {\sc NGC} galaxies. Similar to us, they found that the confidence intervals for the {\sc FDM} mass don't overlap for all galaxies.
Also, scenarios with DM entirely composed of particles with mass in the range $10^{-2} < m_{22} < 10^{2}$ were disfavoured by the analysis of {\sc SPARC} galaxies in \citep{Bar:2021}. \cite{Chan:21} ruled out values of $m_{22}$ in the range $0.14$--$3.14$ from the estimates of the soliton and halo mass of the ESO563-G021 galaxy, which is in agreement with our 95\% credible interval $\mathrm{log}_{10}m_{22}=-2.1^{+0.09}_{-0.10}$. \cite{Robles:19} found it difficult to simultaneously solve the Too-Big-To-Fail problem and explain the observed kinematics of {\sc SPARC} galaxies via {\sc FDM} halo fit based on {\sc FDM}-only simulations \citep{Schive2014, Schive2014a}. \cite{Street:22} did not find any preferable mass for a single-flavour (composed of a single sort of particles) FDM from the combined analysis of data.

Also, \cite{Deng2018, Burkert:20} concluded that it is impossible to explain the observed characteristic radii and core densities of the galaxies within relation \eqref{eq:scaling} with fixed FDM mass. However, there is a concern about the reliability of the analysis in \cite{Deng2018}: the authors implicitly used core sizes and central densities obtained for the Burkert profile and did not analyse the galaxy kinematics directly. 

A big scatter between the values of the FDM mass preferred by individual objects was also found for dwarf spheroidal galaxies (dSphs), which are promising to probe fuzzy dark matter due to their largest mass-to-light ratios compared to other objects. \cite{Gonzales-Morales:2017} asserted an upper 97.5\% bound $m_{22}\lesssim0.4$ from the analysis of the luminosity-averaged velocity dispersion of the stellar subpopulations of classical dSphs. \cite{Chen2017} found that FDM with $m_{22} \sim 1 \text{--} 2$ can fit different kinematic datasets of classical dwarfs. They also concluded that Fornax kinematics requires the ratio of the transition to soliton radii $\alpha\gtrsim2.5$.  

\cite{Pozo:21b} found signatures of transition from dense cores to halos in isolated dwarfs via the Jeans analysis in frames of the FDM model with $m_{22}=1.5$. \cite{Safarzadeh:20} concluded that the FDM model cannot simultaneously explain the density profiles of ultra-faint and luminous dwarf satellites. \cite{Hayashi:21} obtained tight constraints $m_{22}=1.1^{+8.3}_{-0.7}\cdot10^{3}$ on the FDM mass from the kinematic data of ultra-faint dwarf spheroidal galaxy Segue I; however, this result is incompatible with the bounds based on luminous dwarfs, mentioned above. \cite{Dalal:22} suggested the lower bound $m_{22} \gtrsim 4\cdot10^3$ from the stellar kinematics of Segue I and II. A very narrow range around $m_{22} \simeq 10$ was claimed from the existence of a stellar cluster in Eridanus II, which would have to be heated and destroyed by the oscillations of soliton \citep{Marsh:19}. However, \cite{Chiang:21} argued that the core oscillation cannot significantly heat this stellar cluster.  

We find a substantial preference for a soliton density profile from more than half of galaxies in our sample. A related finding was made in \cite{Martino:20}, who reported evidence for the presence of a solitonic core 
in the Milky Way (MW) from the kinematics of stars in the MW bulge. Moreover, such a central soliton may help to form the MW Central Molecular Cloud according to the recent hydrodynamic simulations by \cite{Li:20b}. \citealt{Toguz:22} ruled out mass values in the range $1.6 \leq \log_{10} m_{22} \leq 3.5$ and found a preference for $\log_{10} m_{22} \simeq 1.5$ via the Jeans analysis of the MW nuclear star cluster.

It is worth noting that our results are in agreement with those by \cite{Zoutendijk:21, Zoutendijk:21b}, who argued that the kinematics of ultra-faint dwarf satellites prefer the soliton halo model against {\sc NFW} and Burkert profiles in terms of Bayesian evidences. On the other hand, \cite{Street:22} applied the Bayesian information criterion and found that most of the {\sc SPARC} galaxies favour the Einasto profile against single-flavour {\sc FDM}. However, these results were obtained for different classes of objects, different sets of considered DM profile models (\cite{Zoutendijk:21, Zoutendijk:21b} did not consider the Einasto profile) and with using different {\sc FDM} parametrizations and priors.

Many of the constraints listed above are in tension not only with each other, but also with the Lyman-$\alpha$ lower bounds $m_{22} \sim 10\text{--}100$ \citep{Irsic:17, Nori:19, Rogers:2021}. Moreover, our analysis of many {\sc SPARC} galaxies gives the credible intervals of the {\sc FDM} mass that contradict the Lyman-$\alpha$ constraints. However, the Lyman-$\alpha$ constrains suffer from the degeneracy between the DM properties and the unknown astrophysics at high redshifts.

\section{Discussion}\label{sec:discussion}

Throughout this paper, we use the popular {\sc FDM} density profile proposed by \cite{Schive2014, Schive2014a}. This profile was derived for an isolated relaxed spherically symmetric self-gravitating {\sc FDM} halo without considering any impact from baryonic matter. This might seem like a theoretical oversimplification, but this specific structure of the {\sc FDM} halo, which is composed of a central soliton and an outer envelope, has been found as a common feature in {\sc FDM} simulations under various conditions \cite{Schive2014,Schive2014a,Mocz2019,Schwabe2020, Niemeyer2020, Nori:20, Davies2020, Veltmaat2020, Chan:22}. 

Moreover, it has been shown that solitons survive during {\sc FDM} halo collisions \citep{Veltmaat2016} and during mergers \citep{mina2020solitons}. The latter work also suggests that the scaling relation~Eq.\,\ref{eq:soliton} holds for solitons created during mergers. 

Ultra-light DM particles form a central soliton even in more complex systems. \cite{Schwabe2020} found that inner solitons also form in mixed {\sc FDM} and {\sc CDM} simulations, even if the fraction of {\sc FDM} is only 10\%.

In this work, we also report the deviation from the soliton-halo mass relation predicted in \cite{Schive2014a}. However, different simulations demonstrate a variety in these relations. \cite{mina2020solitons} observed a power-law relation between the soliton and total halo mass, but with the power $5/9$ instead of $1/3$ in Eq.~\ref{eq:soliton-halo-scaling}. A universal relation different from Eq.~\ref{eq:soliton-halo-scaling} was reported in \cite{Mocz2017}. \cite{Nori:20} found that scaling relations are correct only for isolated spherically symmetric and relaxed halos, but are not valid in other cases. The simulations by \cite{Chan:22} established the power $0.515^{+0.111}_{-0.189}$ in the soliton--halo mass scaling relation; in this case, the lower bound is very close to $1/3$ from \cite{Schive2014a} but the mean value is different.

The FDM density profile described in Sec.~\ref{subsec:fdm} was obtained without taking into account the effect of baryonic processes on the DM halo formation. Solitonic cores were also found in the simulations of baryons plus FDM with bayonic feedback \citep{Mocz2019, Veltmaat2020}. According to the simulation in \cite{Mocz2019}, the DM density distribution was mostly unaffected by baryonic processes. Instead, the distribution of baryonic matter in galaxies follows the distribution of FDM halo. On the other hand, \cite{Veltmaat2020} showed that, due to the additional gravitational potential of baryons, FDM forms a denser soliton in the center, but the density profile of the soliton is different than in the case of DM-only simulations Eq.~\ref{eq:soliton}. They also found that the scaling relations Eq.~\ref{eq:scaling}, \ref{eq:soliton-halo-scaling} change due to baryonic feedback. \cite{Davies2020} found that the FDM soliton is denser and smaller in the presence of a supermassive black hole. The difference between the results of the profile between FDM-only and FDM+baryons simulations may give a hint to the solution to the discrepancy between our FDM mass intervals preferred by individual galaxies.

There is a natural question whether some astrophysical objects can mimic the central soliton. Let us consider two such possibilities: giant molecular clouds and central black hole. Our analysis reveals that the MAP values of the central soliton mass are in the range $10^{8}$--$10^{11}\, M_\odot$, which is much larger than the typical mass of molecular clouds, $~ 10^4$--$10^7\, M_\odot$ \citep{Fukui:10}. The obtained {\sc MAP} soliton masses are comparable with the masses of supermassive black holes. However, the obtained maximal a posteriori soliton radii are usually greater than the radii of the innermost rotation curve bins. This can be regarded as an argument against the hypothesis that supermassive black holes mimic the central soliton inferred by our analysis. Also, we perform our fits with spherically symmetric model distributions of dark matter, whereas some {\sc SPARC} galaxies may be explained better by a non-spherical DM distribution \citep{Loizeau:21, Zatrimaylov_2021, Quintana:22}. Non-spherical DM distributions were beyond the scope of the present paper.

Bayesian Inference is a powerful tool but must be used with due caution. For example, independent groups performed a Bayesian analysis on {\sc SPARC} and reported both a non-detection \citep[see, e.g.,][]{Rodrigues:18} and detection \citep[see, e.g.,]{McGaugh:18} of the fundamental acceleration scale (inspired by the Modified Newtonian Dynamics). Furthermore, \cite{Rodrigues:18} found an incompatibility between the confidence intervals of the fundamental acceleration scale for individual galaxies. 

These controversial results caused vigorous discussion about the reliability of Bayesian inference in astrophysics. \cite{Cameron:20} suggested that Bayesian analysis of {\sc SPARC} rotation curves may have pitfalls: i)~the choice of an uninformative prior for the physical parameters; ii)~the credible interval is not the same as the confidence interval and can be defined in different ways. As a particularly illustrative example, \cite{Li:21} showed that a na\"{\i}ve application of Bayesian analysis to the {\sc SPARC} data even allows one to rule out Newtonian gravity: different galaxies prefer incompatible values of the Newtonian gravitational constant $G_N$ if a flat prior for $G_N$ is used. However, this tension disappears if one chooses the log-normal prior. They attribute this `reductio ad absurdum' to the uncertain nature of formal errors. 

Our uninformative prior for $m_{22}$ is motivated by the fact that previous works reported different (and often contradictory) mass ranges. 
We apply the Bayesian evidences calculated for {\sc FDM} with fixed mass on the grid to check the reliability of our $m_{22}$ results. This approach allows one to overcome the shortcoming of the na\"{\i}ve analysis that different definitions of credible intervals may give different results. 

The difficulty of fitting the {\sc FDM} profiles to galaxy data might be that our work (as many others) uses a simple halo model for relaxed spherically symmetric systems. Quite possibly, this is just insufficient to fit a large variety of galaxies. Progress might requires a model of fuzzy dark matter that also takes baryonic matter into account. Na\"{\i}ve application of the Bayesian inference also may have hidden pitfalls. An analysis of ensembles of mock galaxy rotation curves in different dark matter models may be helpful for checking the reliability of the Bayesian analysis and justifying the DM constraints based on the observed kinematics.

\section{Conclusion}\label{sec:conclusions}

In this work, we used Bayesian Inference to find the best dark matter profile for the rotation curves of galaxies in the Spitzer Photometry \& Accurate Rotation Curves (SPARC) database. This approach naturally penalizes models with many parameters and provides us with a posterior parameter distribution. 

We considered four dark matter profiles: cold dark matter (the Navarro--Frenk--White profile, NFW), cold dark matter modified by stellar feedback (cored NFW), empirical cored Burkert profile, and fuzzy dark matter profile. We have taken into account uncertainties in the relations between the soliton mass, the halo mass and the transition radius for the {\sc FDM} model. 

We found that although the {\sc FDM} profile has the largest number of parameters, more than half of the galaxies in the sample substantially prefer this model. Moreover, only 13 out of 138 galaxies disfavor the {\sc FDM} model. We have not found any galaxy that prefers the cusped {\sc NFW} profile.

However, we also found it impossible to satisfactorily fit the rotation curves of all {\sc SPARC} galaxies with a universal value of the {\sc FDM} particle mass. The 95\% credible intervals for the {\sc FDM} particle mass are in tension for different individual galaxies.

The results of our analysis therefore speak both both for and against fuzzy dark matter. 

\section*{Acknowledgements}

We thank Yuri Shtanov, Andrea Ferrara, Pengfei Li and Kirill Zatrymailov for discussions and helpful comments. MKh and AR are grateful to the Armed Forces of Ukraine for protection and security.
This work was supported by the National Research Foundation of Ukraine under Project No.~2020.02/0073. The work of AR was also partially supported by the ICTP through AF-06 and the priority project of National Academy of Sciences of Ukraine No. 0122U002259. MKh acknowledges support from Stiftung Polytechnische Gesellschaft F\"{o}rderung f\"{u}r ein kurzfristiges Projekt zur `Unterst\"{u}tzung gefl\"{u}chteter junger Ukrainischer WissenschaftlerInnen' and FIAS for the hospitality. The calculations presented here were performed on the BITP computer cluster. 

\section*{Data Availability}

The SPARC data underlying this paper are publicly available.  The code that supports the findings of this study will be shared upon a reasonable request to the corresponding author.



\bibliographystyle{mnras}
\bibliography{mylib} 




\appendix

\section{Prior probability density functions}\label{sec:priors}

We assume the lognormal prior for $\Upsilon_\text{disk}$ and $\Upsilon_\text{bulge}$:
\begin{equation}
f(\Upsilon) = \frac{1}{\Upsilon\sigma_{\Upsilon}\sqrt{2\pi}} \exp \left\{- \frac{(\ln\Upsilon -\mu)^2}{2\sigma_{\Upsilon}^2} \right\} \, ,
\end{equation}
where $\mu_\text{disk} = \ln 0.5$ and $ \mu_\text{bulge} = \ln 0.7$. The standard deviation $\sigma = 0.1~\text{dex} = 10^{0.1} \approx 1.26$. \\

Gaussian priors are assumed in the following form: 
\begin{equation}
f(x) = \frac{1}{\sigma\sqrt{2\pi}} \exp \left\{-\frac{(x - \mu_x)^2}{2\sigma^2} \right\} \, ,
\end{equation}
where $\mu_x$ is the mean, and $\sigma$ is the variance.

The flat prior is defined as follows:
\begin{equation}
f(\theta) = 
    \begin{cases}
    \dfrac{1}{(\theta_\text{max} - \theta_\text{min})}, &\text{if} \ \theta_\text{min} \le \theta \le \theta_\text{max} \, , \\
    0, &\text{otherwise} \, , 
    \end{cases}
\end{equation}
where $\theta_\text{min}$ and $\theta_\text{max}$ are the lower and upper bounds, respectively, of the allowed range for each parameter. 

\section{FDM density profile parametrization}
\label{appendix:FDM-parametrization}

The soliton  Eq.\,\ref{eq:soliton} and NFW Eq.\,\ref{eq:nfw-profile} density profiles have the form:
\begin{equation}
    \rho_\text{sol}(r) = \rho_s f_\text{sol} \left( r/r_s \right)\,, \quad  f_\text{sol} (x) = \frac{1} {\left(1 + 0.091 x^2\right)^8}\,,
\end{equation}
\begin{equation}
    \rho_\text{NFW} = \rho_n f_\text{NFW} \left( r/r_n \right)\,, \quad f_\text{NFW}(x) = \frac{1}{x (1+x)^2} \,.
\end{equation}

We define the integrals expressing in dimensionless form the DM mass enclosed within some radius $x$:
\begin{equation}
    I_\text{sol}(x) = \int_{0}^{x}dx'x'^2 f_\text{sol}(x') \,,
\end{equation}
\begin{equation}
    I_\text{NFW}(x_1; x_2) = \text{max} \left[ 0\, , \ \int_{x_1}^{x_2}dx'x'^2 f_\text{NFW}(x') \right] \,.
\end{equation}

In all equations below, we use the dimensionless radius $x = r/r_s$. Now we can write the expression for $v_{200}$ as  
\begin{align}
    v_{200}^2 &= \frac{4\pi G}{r_{200}} \rho_s r_s^3 \Bigl[ I_\text{sol} \left( \text{min} \left[ \alpha\, , c_{200} \right] \right) \nonumber \\ 
    & \quad {} + K \left(\alpha, \beta, c_{200} \right) I_\text{NFW} \left( \alpha/\beta; c_{200} / \beta \right) \Bigr] \,.
\end{align}
where $c_{200} = r_{200}/r_s$, $\beta = r_n/r_s$, $\alpha = r_t/r_s$, and $K \left(\alpha, \beta, c_{200} \right)$ is a factor defined below.  
From the continuity condition for the soliton--halo transition, we have
$$
    \rho_n = \rho_s \frac{f_\text{sol}(\alpha)}{f_\text{NFW}(\alpha/\beta)} \,,
$$
and the factor $K \left( \alpha, \beta, c_{200} \right)$ is given by the relation
$$    
K \left( \alpha, \beta, c_{200} \right) \def \frac{\rho_n r_n^3}{\rho_s r_s^3} =  \frac{\beta^3 f_\text{sol}(\alpha)}{f_\text{NFW}(\alpha/\beta)} \,.
$$
Using the notation introduced above, we obtain an expression for the {\sc FDM} velocity:
\begin{equation}
   \frac{v^2(x)}{v^2_{200}} =  \frac{c_{200}}{x} \frac{I_\text{sol} \left( \text{min}[\alpha, x] \right) + K I_\text{NFW} \left( \alpha/\beta; x / \beta \right)}{I_\text{sol} \left( \text{min}[\alpha, c_{200}] \right) + K I_\text{NFW} \left( \alpha/\beta; c_{200} / \beta \right)} \,.
   \label{eq:FDM-no-scalings}
\end{equation}

We also define an auxiliary function
\begin{equation}
    \mu (x, \alpha, \beta) = I_\text{sol} \left( \text{min} [\alpha, x] \right) + K I_\text{tail} (\alpha/\beta, x/\beta) \,.
\end{equation}
Then the expression for $v_{200}$ becomes
\begin{equation}
    \frac{v^2(x)}{v^2_{200}} = \frac{c_{200}}{x} \frac{\mu(x)}{\mu(c_{200})} \,.
\end{equation}

Now, the {\sc FDM} density profile is parameterized by the set of parameters ($v_{200}$, $c_{200}$, $\alpha$, $\beta$). By taking into account also two scaling relations of the {\sc FDM} model, we obtain parametrization in terms of ($v_{200}$, $m_{22}$, $\alpha$, $\delta$), as stated in Tab.~\ref{tab:model_parameters}. For this purpose, we write the {\sc FDM} scaling relations Eq.\,\ref{eq:scaling} and Eq.\,\ref{eq:soliton-halo-scaling} in terms of new variables. Combining both scaling relations and the definitions of $r_{200}$, $v_{200}$, and $M_{200}$, we get the following equations: 
\begin{equation}
    c_{200} = \frac{0.0537\, m_{22} \delta v_{200}^2 \rho_\text{crit}^{1/3}}{H_0^2}\,,
\label{eq:c200}
\end{equation}
\begin{equation}
    \mu(c_{200}, \alpha, \beta)  = \frac{0.350877\, m_{22}^2 \rho_\text{crit} v_{200}^4}{c_{200} H_0^4}\,.
\label{eq:mu_beta}
\end{equation}

Equation \ref{eq:c200} allows one to express the parameter $c_{200}$ through the new FDM parameters. Equation \ref{eq:mu_beta} should be solved numerically to find the parameter $\beta$. Equation \ref{eq:mu_beta} may not have a solution for some values of the parameters, even within the allowed range. This means that the FDM halo with such parameters does not exist. Therefore, we assign the log-likelihood value $-\infty$ for such a set of parameters.

\section{Posterior probability distribution}\label{sec:posterior}

\begin{figure*}
	\includegraphics[width=1.5\columnwidth]{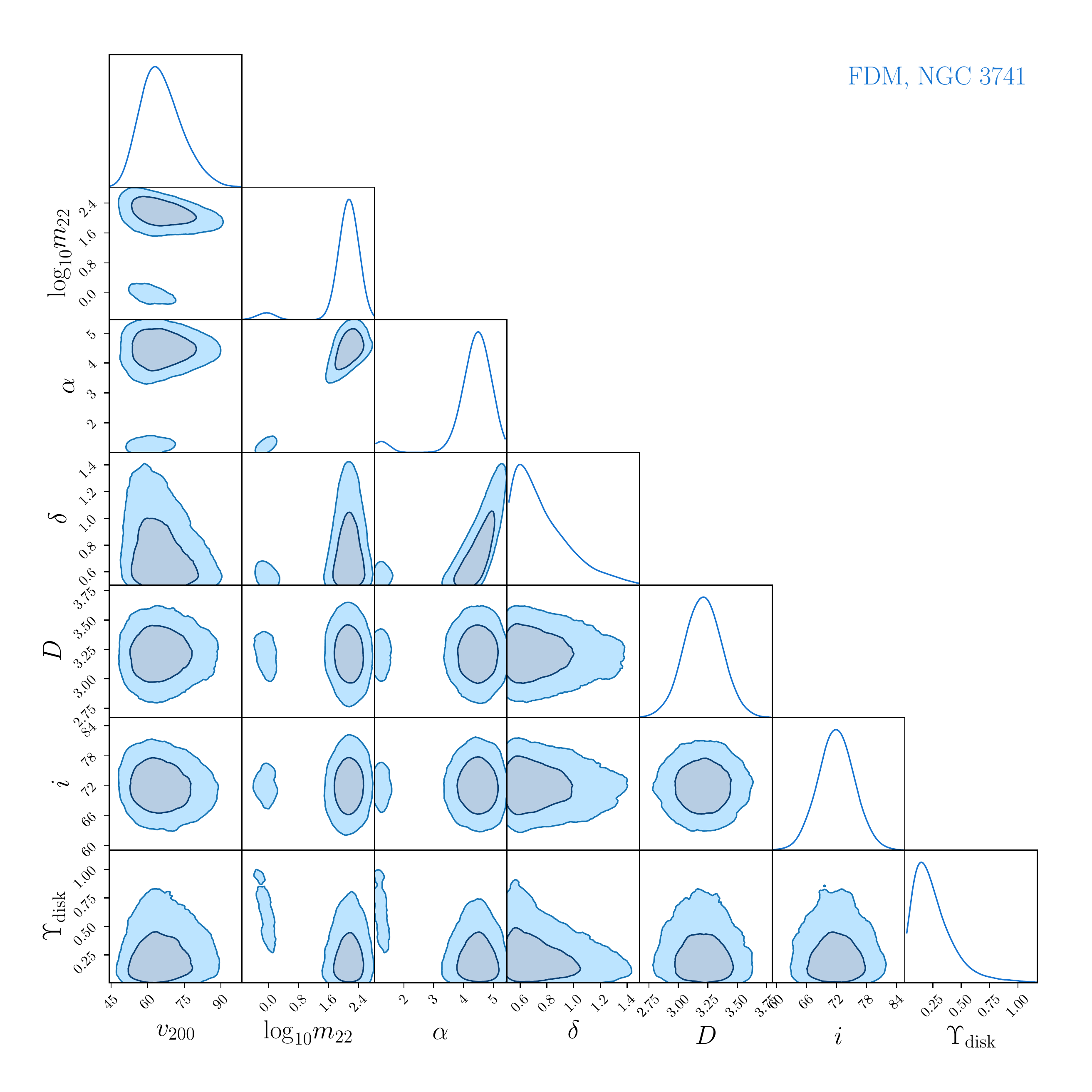}
    \caption{Corner plot showing the parameter constraints obtained from the Bayesian analysis of the rotation curve of NGC 3741: the 95\% contours of the joint 2D marginalised posterior distributions with PDF along the diagonal.
}
    \label{fig:posterior}
\end{figure*}

As an example, the posterior probability distribution of the FDM parameters for the irregular galaxy NGC 3741 is shown in Fig.~\ref{fig:posterior}. The 2D marginalised posterior distributions illustrate the degeneracy between the DM and astrophysical parameters.

\section{Highest-posterior density interval}
\label{appendix:hpd}

We define the highest-posterior density interval  $\left[\theta_1,\theta_2\right]$ for the parameter $\theta$ via the following equations:
\begin{equation}
    P(\theta_1|D) = P(\theta_2|D) \, , \qquad  \int^{\theta_2}_{\theta_1}{P(\theta|D)\mathrm{d}\theta}=a\, ,
\end{equation}
where $P(\theta|D)$ is a posterior probability density for the parameter $\theta$, and $a$ is the fraction of the posterior mass in the interval $\left[\theta_1,\theta_2\right]$.


\bsp	
\label{lastpage}
\end{document}